# Are all citations worth the same? Valuing citations by the value of the citing items*

*Cristiano Giuffrida*
  Department of Computer Science, Vrije Universiteit, Amsterdam
    De Boelelaan 1081A, 1081 HV Amsterdam - THE NETHERLANDS
    giuffrida@cs.vu.nl

*Giovanni Abramo (corresponding author)*
  Laboratory for Studies in Research Evaluation, Institute for System Analysis and Computer Science (IASI-CNR), National Research Council of Italy
    Via dei Taurini 19, 00185 Rome, ITALY
    giovanni.abramo@uniroma2.it

*Ciriaco Andrea D'Angelo*
  Department of Engineering and Management, University of Rome 'Tor Vergata' and Laboratory for Studies in Research Evaluation, Institute for System Analysis and Computer Science (IASI-CNR)
    Via del Politecnico 1, 00133 Rome, ITALY
    dangelo@dii.uniroma2.it

## Abstract

Bibliometricians have long recurred to citation counts to measure the impact of publications on the advancement of science. However, since the earliest days of the field, some scholars have questioned whether all citations should be worth the same, and have gone on to weight them by a variety of factors. However sophisticated the operationalization of the measures, the methodologies used in weighting citations still present limits in their underlying assumptions. This work takes an alternative approach to resolving the underlying problem: the proposal is to value citations by the impact of the citing articles, regardless of the length of their reference list. As well as conceptualizing a new indicator of impact, the work illustrates its application to the 2004-2012 Italian scientific production indexed in the WoS. The proposed impact indicator is highly correlated to the traditional citation count, however the shifts observed between the two measures are frequent and the number of outliers not negligible. Moreover, the new indicator shows greater "sensitivity" when used to identify the highly-cited papers.

## Keywords

*Research evaluation; impact; citing-cited; bibliometrics.*

# 1. Introduction

When the first bibliometricians began exploring the possibilities of metrics in the area of library sciences and established the foundations of what would become scientometrics, they could only draw on rudimentary instruments of computerization. However, they did not lack for pioneering intellectual fervor. The practices that we now take for granted evolved rapidly, following the first timid steps. Those of us who entered this realm of science only this century could already draw on the advantages of much more powerful technologies, and the accumulated knowledge of those who had preceded us. The disadvantage for more creative souls has become that we operate in the context of consolidated scientific paradigms, which channel thought and make the opportunity of any groundbreaking shift difficult and improbable. For many, our destiny is to contribute to an incremental kind of scientific progress, while the heart still aspires to achieve some indelible imprint of creative disruption, giving life to a new paradigm.

What we envy about the fathers of scientometrics is the headiness and excitement of creating an entirely new scientific discipline, and the infinity of research questions offered by an unexplored field. Today it seems almost impossible to raise questions that have not already been addressed by those that preceded us. In the best of cases it seems we can only aspire to offer more complete or somewhat different answers, thanks to the more powerful tools available, or to apply the existing solutions to different contexts.

A question that assails us for some years offers a case in point: Why is it that we generally assign the same value to citations (of the same year and field)? In other words, why are $n$ citations always and in all cases worth more than $n-1$? For bibliometricians, the count of citations received by the knowledge encoded in a publication is a proxy of its future impact on scientific/technical progress, and (once directly or indirectly incorporated in a technology) on economic-social progress. Let us assume that a scientific discovery (encoded in a publication) leads to two other discoveries (publications), of which one provides the basis for a new active substance in a body lotion, and the other the basis for a life-saving pharmaceutical (likely more highly cited than the body lotion publication). Do the two publications citing the first one have the same value? Or is the second one more valuable, given the consequential difference in social impact?

## 1.1 Valuing citations: a brief history

Recalling our opening remarks, it comes as no surprise that others before us would have posed this exact same question. What does seem surprising is that the problem would have been spelled out as long as 40 years ago, specifically by Manfred Kochen (1974), a scholar in information and behavioral sciences, operating in what were still the earliest years of our discipline. Not only did Kochen raise the question, he also suggested a solution: "counting a reference from a more prestigious journal more heavily". Two years later, Pinski and Narin (1976) proposed the first iterative algorithm to operationalize the solution. Cronin (1984) and Davis (2008) also held that the weight of citations should be differentiated to reflect the prestige of citing journals.

With the progress of information technologies, ever more sophisticated algorithms were developed (Liebowitz and Palmer, 1984; Laband and Piette, 1994; Kalaitzidakis, Stengos, and Mamuneas, 2003; Bollen and Van de Sompel, 2006; Kalaitzidakis, Mamuneas, and Stengos, 2011). In 2007 Carl Bergstrom and Jevin West, of the University of Washington, co-founded "The Eigenfactor® Project", aimed at applying network analysis to map the structures of research and assist scholars in navigating the scientific literature. Within The Eigenfactor® Project, and along the scientific paradigm initiated by Kochen (1974), Carl Bergstrom, Jevin West and their colleagues have conceived the Eigenfactor$^{TM}$ score to rate the importance of scientific journals (Bergstrom, 2007; West et al., 2010).[1] In Carl Bergstrom's own words: "This iterative ranking scheme, which we call Eigenfactor, accounts for the fact that a single citation from a high quality journal may be more valuable than multiple citations from peripheral publications" (Bergstrom, 2007). The Eigenfactor$^{TM}$ score then embeds weighted citations.

Several years later, the SCImago research group developed its own iterative algorithm, based on citations weighted by the visibility of the citing journal. The algorithm was applied to rank journals, in the form of the SCImago Journal Rank or SJR (González-Pereira, Guerrero-Bote, & Moya-Anegón, 2010). Soon after this, Guerrero-Bote & Moya-Anegón (2012) developed a more sophisticated variant of the SJR, known as SJR2.

The "higher order" evaluation method, originally conceived by Pinski and Narin (1976) for ranking journals, has more recently been applied for a series of purposes, thus ranking: individual publications (Chen et al., 2007; Walker et al., 2007; Ma et al., 2008; Li & Willett, 2009; Yan & Ding, 2010; Su et al., 2011); authors (Fiala et al., 2008; Ding et al., 2009; Radicchi et al., 2009; Ding, 2011; Fiala, 2011; Yan and Ding, 2011; Fiala, 2012b; Fiala, 2013a; Nykl et al., 2014); departments and institutions (Fiala, 2013b; Fiala, 2014; Yan, 2014); countries (Ma et al., 2008; Fiala, 2012a); an integration of publications, journals, and authors (Yan et al., 2011); a mixture of the preceding entities (West et al., 2013).[2] Over the course of decades, the original concept at the basis of the weighted citation count rating for journals has thus gradually been translated to rating authors, institutions, countries and more. In West et al.'s (2013) own words, regarding the Eigenfactor$^{TM}$ score adapted to rate authors: "The Eigenfactor$^{TM}$ score can be viewed as a form of weighted citation count where the weights reflect the prestige of the citing authors".

**1.2 Towards a new paradigm**

In 2004 the current authors but Cristiano Giuffrida co-founded the National Research Council of Italy and University of Rome "Tor Vergata" joint "Laboratory for Studies on Research and Technology Transfer", since renamed the "Laboratory for Studies in Research Evaluation". Our aim was mainly to provide policy makers and the management of research organizations with diagnostic tools and performance indicators, for assessment of scientific strengths and weaknesses at the national and institutional levels. We have now spent a number of years applying our citation-based indicator, Fractional Scientific Strength (FSS) (Abramo & D'Angelo, 2014), to measure the scientific performance of individuals and organizations at field and discipline levels. FSS is a size-independent citation-based indicator based on the ratio of outcome to input, which differs substantially from widely used "per publication" citation-based indicators (Abramo &

---

[1] The Eigenfactor scores and its variant, Article Influence, are available online at http://www.eigenfactor.org, without cost, last accessed on 15 Feb. 2019.
[2] For an in depth review of the work that has been done in the field, we refer the reader to Waltman and Yan (2014).

D'Angelo, 2016a; 2016b), such as the well known Mean Normalized Citation Score (MNCS) (Waltman et al., 2011). Given our professional aims and activities, it was no wonder that the question as to whether citations have different values soon sprang to mind.

What then is the reason that we now wish to address such a question in this manuscript, particularly since the answer and its various operationalizations have already been available in the literature, in some cases for many years? The problem is that we are not conceptually satisfied with the solutions provided, for two fundamental reasons. We return to the above case of the scientific discovery (publication A) which leads to two further discoveries (publications B and C), where B gives rise to a new active substance in a body lotion and C gives rise to a life-saving pharmaceutical, and then the question as to whether the citation by C should have greater weight, given the different social impact that it originates. Since the time of Kochen (1974) and until Chen et al. (2007), the paradigm guiding the development of iterative algorithms provides that: i) a weight must be assigned to the citations; ii) the weight depends on the influence of the citing journal.

Instead, we think that it would be more correct to value a citation in function of the impact of the citing article, rather than the journal (it is the greater number of citations which C would presumably receive that reflects the differential impact, rather than the prestige of the journal in which it is published). In support of our objection, we recall that weighting a citation by the influence of the citing journal is in conflict with what we know about measuring the impact of a publication. We refer the reader to a recent work by Abramo (2018) on this specific issue. Suffice here to say that for a publication to have a scholarly impact, it has to be used by other scientists: no use, no impact. Consequently, citation is the natural indicator of impact, as it certifies the use of the cited publication towards the scientific advancement encoded in the citing publication.

No other bibliometric indicator certifies use better than citations.[3] The journal impact, in particular reflects the distribution of citations of all hosted publications, not the individual ones. In the 2013 San Francisco Declaration on Research Assessment (DORA) recommends[4] against using the journal impact (IF) as a substitute measure of the impact of individual research articles, and states that such practices create biases and inaccuracies when appraising scientific research. Exceptions may be considered only for very young articles.

The combination of journal with citation metrics has in fact been recommended only for zero or one-year citation windows (Levitt & Thelwall, 2011), and for a two-year window in the case of papers in mathematics (with weaker justification in biology and earth sciences), because of the characteristic inertia of these disciplines regarding the early stages of accruing citations. Confirming Levitt and Thelwall (2011), in the social sciences, the IF is seen to improve the correlation between predicted and actual ranks by citation only when applied in the "zero" year of publication and up to one year afterwards (Stern, 2014). For citation time windows above two years, citation shows a stronger predictive power than the IF alone (Abramo, D'Angelo & Di Costa, 2010), or an *a priori* combination of citation and IF (Abramo & D'Angelo, 2016c). The appropriate combinations of citation and IF per scientific field and citation time window have been further provided by Abramo, D'Angelo and Felici (2019). Finally, we must also recall that some citations inevitably originate from publications that remain unrated in terms of IF or the like.

---

[3] The limits of this statement and a discussion on social constructivism vs the normative theory of citing can be found in Abramo (2018).
[4] http://www.ascb.org/dora/, last accessed on 15 Feb. 2019.

For this, we would prefer to value a citation by the field-normalized citations accumulated by the citing article. Put simply, two citing articles with different field-normalized citations would determine a measurement of (predicted) impact of the cited article different from the simple tally of citations (i.e. 2), and likely different from that derived from IF-weighted citations (although a certain correlation is to be expected).

By the same reasoning, we have problems with the adaptation of the Eigenfactor$^{TM}$ score to rank authors, institutions, and nations, whereby the weights reflect respectively the prestige of the citing authors, institutions, and nations. In our view, the weights should reflect the "importance" of the citing articles rather than that of the citing authors, institutions, and nations. Because it is the final impact of the new knowledge produced that scientometricians want to measure, and that determines the scientific "importance" of authors, institutions, and nations. Although a certain correlation is to be expected in the final outcomes, there is a nuanced yet substantial conceptual difference between the two approaches. The impact of the new knowledge produced should be evaluated through the citation network of the publication at stake, where the nodes are represented by the citing publications only, and not by other "surrogate entities".

Chen et al. (2007) were the first to propose a PageRank-inspired method for analyzing publication citation networks. Chen et al.'s proposed indicator to measure a publication's impact is based on and reflects the following assumptions: i) being cited by higher-cited papers contributes more to final impact than being cited by lower-cited papers; (ii) being cited by a paper that itself has few references gives a larger contribution to impact than being cited by a paper with a higher number of references. Walker et al. (2007) followed suit, trying to correct the tendency of Chen et al.'s method to favour older publications. Yan and Ding (2010) give more weight to articles that are cited immediately than to those being cited at a later date. Su et al. (2011) proposed a solution suitable for cases where there are missing papers in the database citing network. Other scholars measured the impact of publications through their citation network, in different contexts and disciplines (Ma et al., 2008; Li and Willett, 2009).

**1.3 Our approach**

The methodology that we propose in the current work departs from the above ones in two ways. First, the length of the reference list should not affect the measurement of impact. From an economic perspective, the value of a citation is independent of the number of publications (whether high or low) the citing article cites. Second and more important, according to the above PageRank-inspired methods, all others equal, the impact of a publication gathering one citation only can be higher than that of a publication gathering two or more citations.

Our stance instead is that the value of the citing article should not be entirely transferred to the cited publication, but only in such measure that no individual citation (even one from a highly cited paper) be allowed to be worth above two, which is the value of two citations from uncited publications.[5] In fact, we hold that however high the impact of a citing article may be, its impact should not be directly transferred to the cited publication, as happens when citations are weighted accordingly. The underlying rationale is that many cited references will offer no direct or relevant contribution to the new knowledge encoded in the citing article (towards the new publication concerning the life-saving pharmaceutical, for example). Those very few references that do make substantial contribution are themselves likely to be cited in manner proportional to their relevance. In other words, if a publication is a star it does not need to leech a citing publication to get its light. Vice versa, if a publication is not a star, as is more often the case, the bibliometrician should avoid the risk of rating it as a star just because it has been cited by a star publication.

---
[5] We are implicitly assuming that a citation cannot be worth less than one.

We are aware though that the idea that a citation may be worth more than another but not very much more, might appear too conservative to many. We propose then two formulas to value citations: one based on our convention that no individual citation (even one from a highly cited paper) be allowed to be worth above two, which is the value of two citations from uncited publications; and the other releasing the above constraint. Both formulas allow anyway for a wide continuum on which the conversion between citations (i.e. differential weighting) can be carried out.

In this work we provide conceptual and operative illustrations of our approach to valuing citations, which we apply to the Italian WoS-indexed publications from the 2004-2012 period. We then compare the results with those obtained from the application of the traditional method. Sections 2 and 3 present the conceptual framework and the operative method of measurement. Sections 4 and 5 illustrate the results from the application and comparisons. Section 6 provides our concluding remarks.

## 2. Conceptual framework

Traditionally, the measurement of the impact of a publication requires counting the citations it receives and then standardizing them in function of both the reference scientific domain and the "age" of the publication. However this procedure ignores the fact that scholarly references join together in a vast network of citations, in which each citing publication is itself more or less cited. We take the case indicated in Figure 1, where two publications ($\alpha$ and $\gamma$) are issued on the same date and belong to the same scientific domain. Within a given citation time window, both of these receive three citations: $\alpha$ from publications A, B and C; $\gamma$ from D, E and F (Level 1). At the next higher level (Level 2), within the same citation window, we see that A, B and C have in turn received only one citation and that all of these are from the same publication (a), while D, E and F are respectively cited by three (a; b; c), three (D; d; e) and one publication (f). We can also see that publication D belongs both to citing Level 1, since it cites $\gamma$, and to citing Level 2, since it cites E, which in turn cites $\gamma$.

*Figure 1: Network of citations, an example*

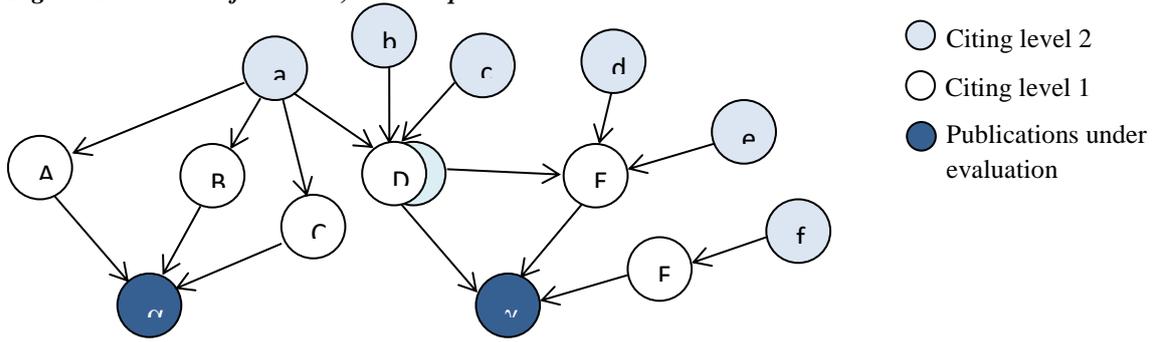

Figure 2 shows the partition of the overall network, concerning two and only two levels. Limiting ourselves to observing just one level (Level 1) we can only note that α and γ receive the same number of citations, and consequently, we would conclude that they show no difference in impact. However, in broadening the observation to the higher level citational network (Level 2), we observe that while articles (A, B and C) citing α have had an impact limited to a single work (a), those (D, E and F) that cited γ have had an impact on seven different works. Note that D belongs to both the level 1 (it cites γ) and the level 2 (it cites E which cites γ). Publications a,b,…f could obviously in turn be cited by other publications at Level 3, and so on. Clearly then, the more levels of the network we are able to observe, the richer would be the dataset on which we could base the evaluation of the impact of publications α and γ, and the more precise would be the relative measure. On the other hand, the more we ascend the chain of levels, the more we restrict the time window for the citing publications (in turn rendering the use of early citations as proxy of impact ever less accurate), and the greater become the computational complexities. Also, and much more important, in practical applications the decision-maker typically requires evaluation of performance in the near term (timeliness in research assessment), meaning with quite short citation time windows: conditions under which analysis above Level 2 would lack precision and be poorly representative. For this reason, as well as for simplicity in demonstration in the current work, we will limit our observations to what happens in changing over from the traditional measure of publication impact (simple citations count – Level 1), to a measure that considers the citations of the citing publications (Level 2), all within the same citation time window.

## 3. Method

Traditionally, the impact of a publication is measured by tallying citations. Since citation behavior varies across fields, accumulated citations are evidently a function of the "quality" of the cited publication and of citation time window, but also of the field to which the cited publication belongs. To avoid distortions in the comparisons, bibliometricians normalize the citations by a scaling factor. We adopt the average of the distribution of citations received for all cited publications of the same year and subject category ($c_{exp}$).[6] Carrying out this normalization, we obtain the field-normalized citation score:

---

[6] Abramo, Cicero and D'Angelo (2012) have demonstrated that this is the most effective scaling factor.

$$C = \frac{N}{c_{exp}}$$

[1]

where *N* is the number of citing publications.

Referring to Figure 1, the traditional method of measuring the impact of α and γ would observe only Level 1, and conclude that the publication with highest value of $C$ is that with the greatest impact. In this case, having both received three citations and being works published on the same date and belonging to the same scientific domain (same $c_{exp}$), α and γ would present the same impact.

The method we propose would instead extend the observation to the next levels (in the case of Figure 1, to Level 2). What we wish to do is take account of the fact that each of the *N* publications citing α and γ in turn receives a number $c_i$ of citations, on the basis of which we could differentiate the contribution of each of these in determining the impact of α and γ. The point becomes how to differentiate the contribution. In determining this, our underlying rationale is that the differentiated contribution of each citing publication must be proportional to its own impact, but at the same time neither penalize the cited one (in the case that the citing publication is not cited itself), nor excessively reward it (in the case that the citing publication is very highly cited): both of these being cases that would arise through weighting (which implies a multiplication). Therefore, the method of valuing the citing publications must respect the condition that two citing publications (even if uncited) cannot count less than one (even if highly cited). Furthermore, among the various distributions that are characterized by the high skewness typical of citation counts, such as power law, lognormal, etc. we assume an exponential model.

The new indicator proposed to account for the different contribution of citing publications is $C_v^*$, which we measure as follows:

$$C_v^* = N + \sum_{i=1}^{N} e^{\beta \cdot f(c_i)}$$

[2]

with

$$f(c_i) = 1 - \frac{C_{i\_max}}{c_i}$$

[3]

where *N* is the number of citing publications; $c_i$ is the number of citations received by the citing publication *i*; $C_{i\_max}$ ($\neq 0$) is the maximum of the distribution of citations received by all cited publications of the same year and subject category (SC). The new indicator can assume values between *N* (in the case that none of the citing publications is in turn cited) and *2N* (in the case that all the citing publications are the highest cited among those of the same year and SC).

The parameter *β* must be inevitably determined on the basis of a convention. In our case, such convention can only be based on empirical data available to us, i.e. Italian publications indexed in WoS over the period 2004-2012.[7] Given the citational distribution of such publications for each year and SC, we extract the median and maximum of the said distributions and calculate their ratio. The average of such ratios results as 0.05, from which we impose that: when the number of citations received by the citing publication equals 5% of the maximum of the relative reference distribution, the weight of the citing publication must be 1.5. In other words, we assume that the citing publications with a "median" impact (as measured by the traditional approach) have a weight 50% higher than that of an uncited citing publication.

---
[7] If world baselines were available to us, we would certainly use them.

From this it follows that:[8]

$$\beta = -\frac{1}{19}\ln\frac{1}{2} = 0.03648 \approx \frac{1}{10 \cdot e}$$

(e = Euler's constant, 2.718)

Figure 2 shows the empirical curve $C_v^*$ in function of the "gain", or $\frac{c_i}{C_{i\_max}}$.

The value of $C_v^*$ calculated using [2] is obviously also influenced by the field and citation time window, exactly as for $N$. Therefore we must again carry out the rescaling of $C_v^*$ with respect to the expected value, referred to the distribution of publications for the same year and SC. We thus arrive at the field-normalized indicator $C_v$:

$$C_v = \frac{C_v^*}{C_{v_{exp}}^*}$$

[4]

in which $C_{v_{exp}}^*$ is given by the average of the values of $C_v^*$ referred to all the publications of the same year and SC.[9]

In order to exemplify the method and offer a concrete, small-scale illustration of its application, we refer the reader to Figure 1 representing a whole citation network related to a given subject category. The graph shows 14 publications (nodes) and 16 citations (edges), so that $c_{exp} = 1.143$ (16/14) and $C_{i\_max} = 3$ (max number of incoming edges). Since both α and γ receive 3 citations each, their impact would be the same by the traditional approach, $C = 2.625$ (3/1.143). Indeed A, B, C and F receive one citation each, so that for them $f(c_i) = -2$, while E and D receive three citations with $f(c_i) = 0$. Applying equation [2] to α and γ, we register respectively, $C_v^* = 5.789$ and $C_v^* = 5.930$. According to the network representation in Figure 1, publications at citing level 2 are not cited ($c_i = 0 \Rightarrow f(c_i) = -\infty$). Therefore, for all other publications but E, $C_v^* = N$. Conversely, for E $C_v^* = 4$, summing up the two citations from d and e and the one from D with weight 2, being cited in turn 3 times, namely the maximum. The average of $C_v^*$, i.e. $C_{v_{exp}}^*$, is equal to 2.840, so that $C_v = 2.038$ for α, and $C_v = 2.088$ for γ.

The extension of the method to other citing levels beyond 2 is conceptually straightforward, although operationally more complicated.

*Figure 2: Empirical distribution of $C_v^*$ vs the "gain", i.e. $\frac{c_i}{C_{i\_max}}$*

---

[8] Note that, in choosing to average median to maximum citations ratios for having one and only one β for all subject categories, we are assuming that the contribution of citing publications must be field independent. Furthermore, the maximum value of the citation distributions is an outlier by definition, so it seems of little use to recur to different βs for different subject categories, given the instability of the denominator in such ratios.

[9] For more effective scaling of $C_{v_{exp}}^*$ (and also for scaling of $c_{exp}$) the calculation excludes the publications with nil values of $C_v^*$. In addition, while calculation of $c_{exp}$ considers all world-wide publications, for $C_{v_{exp}}^*$ the world-wide distribution of $C_v^*$ is unavailable. In the elaborations that follow, this means that for $C_{v_{exp}}^*$ we are obligated to use the distributions referred to Italian scientific production. We note that the aim of the current work is to illustrate the conceptual proposal and the method of measurement: the results from applying the method can vary according to the conventions adopted and the specific context.

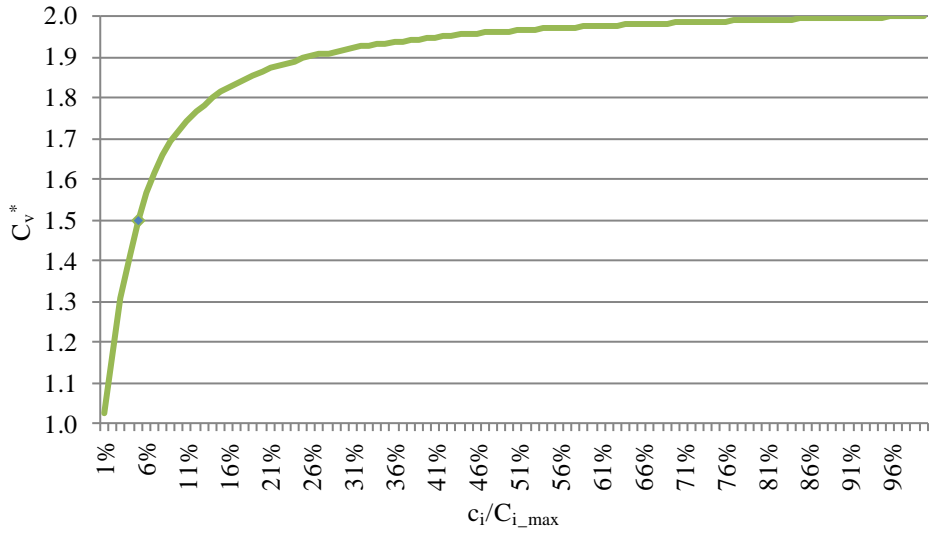

The proposed model embeds three arbitrary conventions: i) the cap value of a single citation; ii) the exponential function to allow for a wide continuum on which the conversion between citations (i.e. differential weighting) can be carried out; and iii) the determination of the parameter $β$, which sets the "conversion rate". Such conventions reflect our personal thinking, but can be modified to accommodate different perspectives. In particular, equations [2] can be replaced by the following:

$$C_v^* = \sum_{i=1}^{N} (1 + \frac{c_i}{c_{i\_exp}})^γ$$

[5]

where $c_{i\_exp}$ is the mean or the median of the citation distribution (in place of the maximum which usually is an outlier, not representative of the characteristics of a citation distribution as a whole); and γ ranging between 0 and 1, allows to accommodate different perspectives: in the extreme case in which γ is set to zero, each citation is worth 1 (the traditional approach, in which all citations are counted equally). The other extreme is γ = 1, which is similar to the PageRank approach. For any γ > 0 there is no upper bound for the weight of a citation, meaning that this model can in no case comply with the convention adopted in our model.

## 4. Application to 2004-2012 Italian publications

We apply the new $C_v$ indicator to analyze the dataset consisting of the WoS Italian Citation Report, extracted from the 2004-2012 WoS Core Collection by imposing the word "Italy" in searching the authors' affiliations. Citations are observed as of 31/12/2014, giving a citation time window broad enough to assure robustness of citations as a proxy of impact (Abramo, Cicero, & D'Angelo 2011). Again for reasons of significance, in terms of citations as a proxy of impact, we exclude the publications in the SCs pertaining to "Art and Humanities". Finally, the analysis excludes non-cited publications, since these present nil impact independent of the indicator used. The final dataset consists of 458,658 publications. For each of these we calculate the proposed indicator as defined in [4] and compare the results with those from the traditional indicator as defined in [1]. In the following we illustrate some examples of this comparison, for purposes of:
- measuring the level of convergence between the distributions deriving from the two indicators, and the extent and kinds of shifts in specific situations;
- illustrating some statistical characteristics of the indicators – in particular, through the variation coefficient, we examine their capacity to capture significant differences in impact between publications;
- comparing the right tails of the two distributions, to verify whether the highly-cited publications under one indicator remain highly-cited under the second indicator.

We begin with two publications in Hematology from year 2012: Table 1 presents the relative bibliographic references and the values for our two indicators. The first publication received 12 citations,[10] compared to an average of 11.87 for publications in the same year and SC. Given this, the observed value of $C$ is just more than one (1.012). The second publication received less citations (10), from which we have a value of 0.842 for $C$, however the citing publications are in turn more cited than those citing the other publication: the values for the $C_v$ indicator are thus observed as 0.743 for the first publication compared to 0.919 for the second. In substance the first publication shows higher impact than the second if measured by $C$, lower if measured by $C_v$. Underlying this observation are the facts that the 12 publications citing the first one have in turn accumulated 19 citations, while the 10 works citing the second have gathered a full 65 citations.

*Table 1: Bibliographic and citational references for two publications in Hematology, 2012*

| WoS code | 309242000007 | 309011200016 |
|---|---|---|
| Author(s) | Montalban et al., 2012 | Vago et al., 2012 |
| Title | Risk stratification for Splenic Marginal Zone Lymphoma based on haemoglobin concentration… | T-cell suicide gene therapy prompts thymic renewal in adults … |
| Source | British Journal of Haematology | Blood |
| DOI | 10.1111/bjh.12011 | 10.1182/blood-2012-01-405670 |
| N | 12 | 10 |
| $C$ | 1.011 | 0.842 |
| $C_v$ | 0.743 | 0.919 |

As a further example we consider four publications from 2011 in Engineering, mechanical, indicated in Table 2. Having all received the same number of citations (21), they all show an identical value of $C$, at 2.974: in other words almost three times the world-wide average for 2011 publications in Engineering, mechanical. Still, considering the impact of the citing articles, we observe important differences: in the last column of the table we see that the values of $C_v$ vary from a minimum of 2.413 for the first article to a maximum of 4.227 for the last. Once again the variance in $C_v$ is explained by the variance of the citations received for the citing articles. The first publication is in fact cited by publications that in turn receive only 14 citations, against 50 for the second article and 215 and 194 for the third and fourth, respectively. However, we also see that in spite of the differential between the third and fourth, the latter still exceeds the third in terms of $C_v$, given the normalizations involved in the indicator (concerning year and SC of each citing work).

---

[10] i.e. observed as of 31/12/2014

*Table 2: Bibliographic and citational references for four publications in Engineering, mechanical (2011), each with 21 citations*

| WoS code | Authors | Title | Source | $C_v$ |
|---|---|---|---|---|
| 299562000004 | Poussot-Vassal et al., 2011 | Vehicle dynamic stability improvements through … | Vehicle System Dynamics | 2,413 |
| 291316000024 | Anzalone et al., 2011 | Advanced Residual Stress Analysis and FEM … | J Micro Electromechanical Systems | 2,927 |
| 285726600012 | Angrisani et al., 2011 | Experimental investigation to optimise a desiccant… | Applied Thermal Engineering | 3,951 |
| 284970700015 | Ferreira et al., 2011 | Analysis of thick isotropic and cross-ply laminated … | Journal of Sound and Vibration | 4,227 |

Figure 3 shows the dispersion of values for the two indicators $C$ and $C_v$, for all Italian publications in Transportation in the year 2012. As expected, the correlation between the two indicators is clearly high: the R-squared regression is greater than 0.92, although we can see that some publications depart from the plot, particularly in the central part of the diagram.

*Figure 3: Dispersion of $C$ vs $C_v$ for Italian publications in Transportation, 2012*

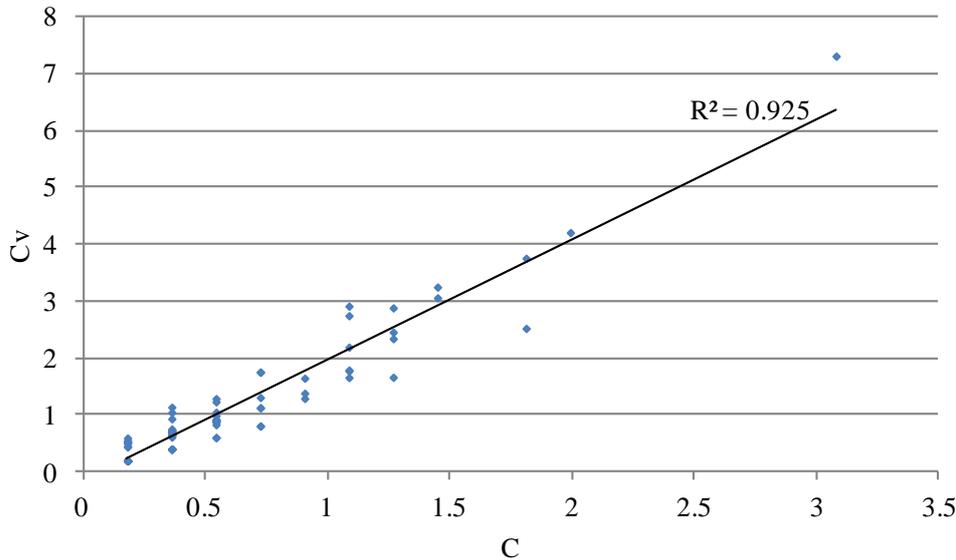

Even in distributions with greater linear fitting of data, we still observe the presence of outliers. The case of the 2007 publications in Economics (Figure 4) offers an example, where we see that the R-squared for the $C$ versus $C_v$ regression is nearly 0.99. Yet, as in other cases, a number of publications still deviate significantly from the plotted line.

*Figure 4: Dispersion of C vs $C_v$ for Italian publications in Economics, 2007*

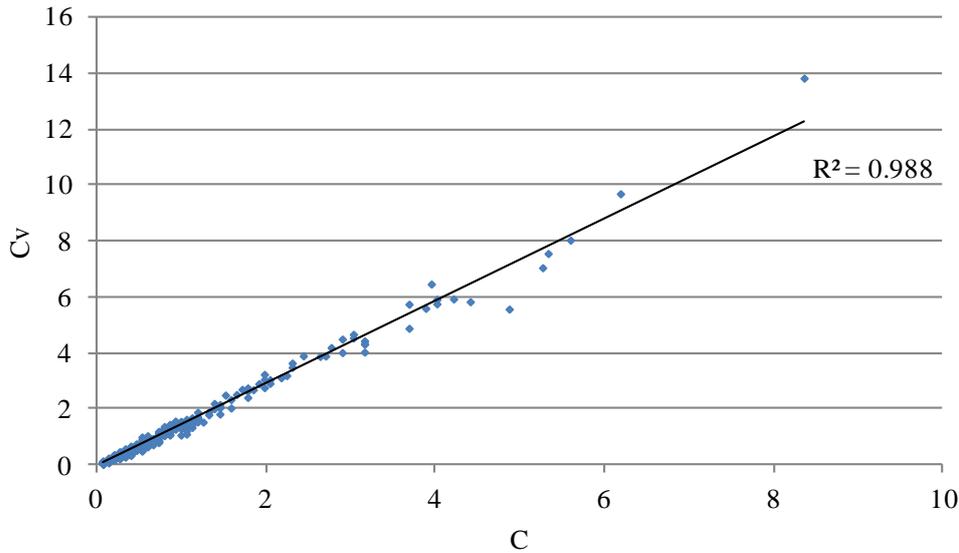

We have repeated the regression analysis for the dispersion of data for all SCs in all years. For reasons of space we show further examples of only 11 SCs: as a criterion for selection, the examples we show are the SCs with the largest number of publications, from each macro-area.[11] The results of the regression analyses, divided by year, are shown in Table 3. In no case do we observe an R-squared less than 0.9, confirming the high convergence of the measures using the two indicators.

*Table 3: R-squared linear regression of C vs $C_v$ for the largest subject categories in each WoS macro-area*

| Year | Mathematics | Astronomy & Astrophysics | Chemistry, Multidisciplinary | Environmental Sciences | Biochemistry & Molecular Biology | Oncology | Surgery | Psychology, Experimental | Engineering, Electrical & Electronic | Economics | Political Science |
|---|---|---|---|---|---|---|---|---|---|---|---|
| 2004 | 0.991 | 0.998 | 0.998 | 0.999 | 0.998 | 0.999 | 0.997 | 0.998 | 0.994 | 0.997 | 0.989 |
| 2005 | 0.993 | 0.997 | 0.998 | 0.998 | 0.999 | 0.999 | 0.998 | 0.999 | 0.994 | 0.994 | 0.982 |
| 2006 | 0.995 | 0.998 | 0.999 | 0.996 | 0.998 | 0.998 | 0.995 | 0.996 | 0.989 | 0.996 | 0.994 |
| 2007 | 0.993 | 0.997 | 0.999 | 0.997 | 0.998 | 0.998 | 0.995 | 0.997 | 0.992 | 0.988 | 0.992 |
| 2008 | 0.987 | 0.997 | 0.998 | 0.995 | 0.998 | 0.999 | 0.995 | 0.996 | 0.990 | 0.993 | 0.977 |
| 2009 | 0.983 | 0.998 | 0.996 | 0.996 | 0.998 | 0.998 | 0.996 | 0.992 | 0.989 | 0.993 | 0.991 |
| 2010 | 0.979 | 0.998 | 0.996 | 0.996 | 0.997 | 0.998 | 0.993 | 0.993 | 0.986 | 0.987 | 0.984 |
| 2011 | 0.969 | 0.997 | 0.995 | 0.996 | 0.997 | 0.999 | 0.991 | 0.988 | 0.983 | 0.962 | 0.942 |
| 2012 | 0.986 | 1.000 | 0.995 | 0.987 | 0.993 | 0.998 | 0.983 | 0.987 | 0.974 | 0.968 | 0.989 |

The comparison between the variation coefficients for the distributions of the indicators is a valid aid in understanding which is better able at capturing significant differences between observations. As an example, for all Italian publications in 2008, Figure 5 shows the trends of the variation coefficients for distributions of $C$ versus $C_v$ in the 22 SCs where the number of publications is not less than 600. The variation coefficient for $C$ is greater than for $C_v$ in only six cases: in Oncology (DM); Cardiac & Cardiovascular Systems (DQ); Endocrinology & Metabolism (IA); Mathematics, Applied (PN); Clinical Neurology (RT); Physics, Multidisciplinary (UI). In the other 16 SCs the opposite occurs: the indicator $C_v$ demonstrates greater variability than $C$.

---

[11] Our assignment of SCs to macro-areas (Mathematics; Physics; Chemistry; Earth and Space Sciences; Biology; Biomedical Research; Psychology; Clinical Medicine; Engineering; Economics; Law, political and social sciences) follows a pattern previously published in the ISI Journal Citation Reports website, although this information is no longer available through the Clarivate web portal. There are no cases where an SC is assigned to more than one macro-area.

Considering all the years and extending the analysis to all SCs with at least 30 publications in each year, the share of SCs where variability of $C_v$ is greater than that for $C$ is never less than 75%, and this share tends to increase with decreasing citation time window, as seen in Figure 6. These results indicate that $C_v$ serves better in discriminating the differences in impact between publications, and that this greater capacity increases as the citation time window decreases.

Figure 7 shows the dispersion for the two indicators ($C$ and $C_v$) for a random sample of 10,000 publications taken from the dataset. The right diagram refers to the top 5% of publications by $C$, and shows a linear fitting which is still significant, although slightly weaker, with an R-squared of 0.916 compared to 0.947 for the entire sample.

*Figure 5: Variation coefficients of distributions for $C$ vs $C_v$ for 2008 Italian publications, in subject categories with at least 600 publications*

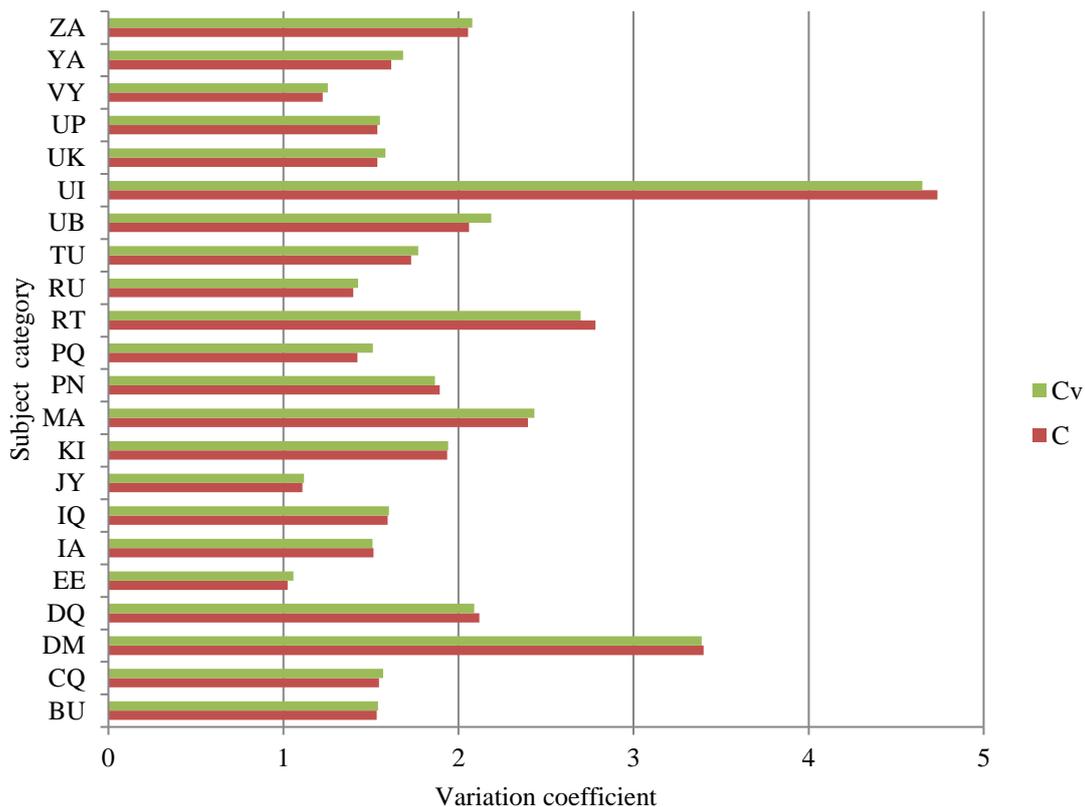

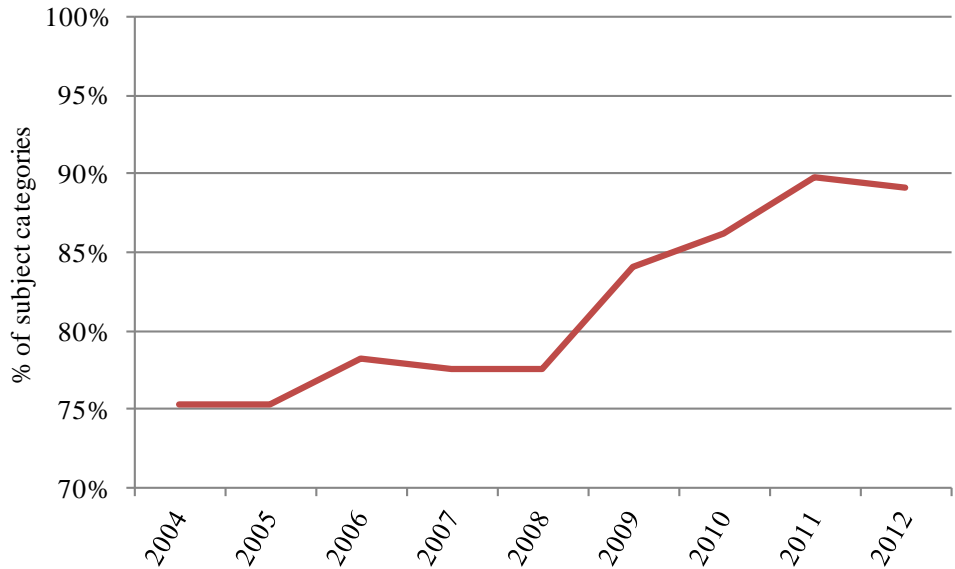

*Figure 6: Percentage of SCs with variation coefficient for distributions of $C_v$ greater than that for C*

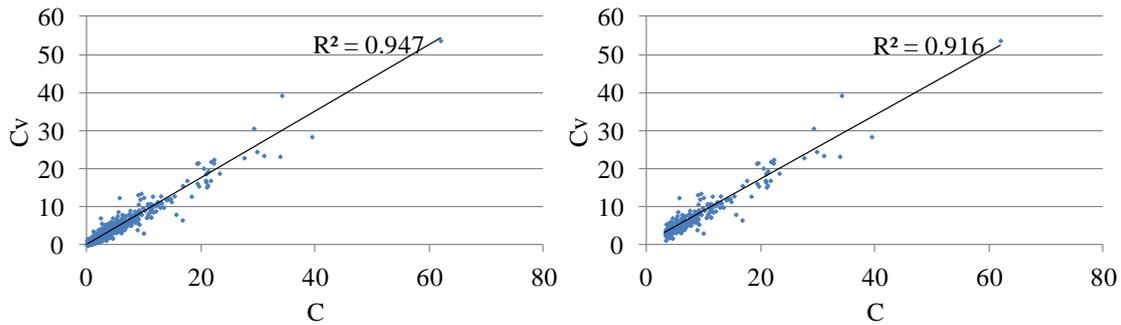

*Figure 7: Dispersion of C vs $C_v$ for a sample of 10,000 2004-2012 publications*

The analysis of the so-called highly cited articles (HCAs) offers important clues, since it deals with the outliers of the citational distributions, meaning publications of very high impact, therefore generally objects of great interest. In particular, we can quantify the cases of publications that are top ranked on the basis of one indicator but not "top" under the other indicator. In general, defining highly-cited publications as those above the 90th percentile for the reference indicator, we observe that 13.5% defined as such for the $C$ distribution are not top for $C_v$, and vice versa, 13.5% of those that are top for $C_v$ are not so for $C$. Restricting the analysis to the top 5% of the distribution, the latter percentage rises to 15.6%, and finally considering the top 1%, to 17.0%. Apart from the generally high correlation between the two distributions, the right tail thus seems to be more influenced by the change in indicator.[12]

We also ask whether in identifying the works of greatest impact, the two indicators give rise to polarized results on certain subject categories. Once we have defined the threshold value for qualifying the top works, for example at 90%, the expected percentage of highly-cited papers would be around 10% in each SC, for both indicators. The questions are thus whether the two indicators are capable of respecting this share, and which shows the least fluctuations. Figure 8 presents the results of such analyses, diagramming the incidence of -publications in the SCs (with at least 30 publications) for 2008. We observe that the percentages fluctuate around the benchmark value (10%), but that the fluctuations are clearly greater for indicator $C$ than for $C_v$.

---

[12] As previously noted, in calculating $c_{exp}$ we can consider all world-wide publications, while for $C^*_{v_{exp}}$ the lack of world-wide distributions of $C^*_v$ requires the resort to distributions referring only to the population under examination (in our case, Italian scientific production).

*Figure 8: Share of top 10% cited 2008 publications among subject categories*

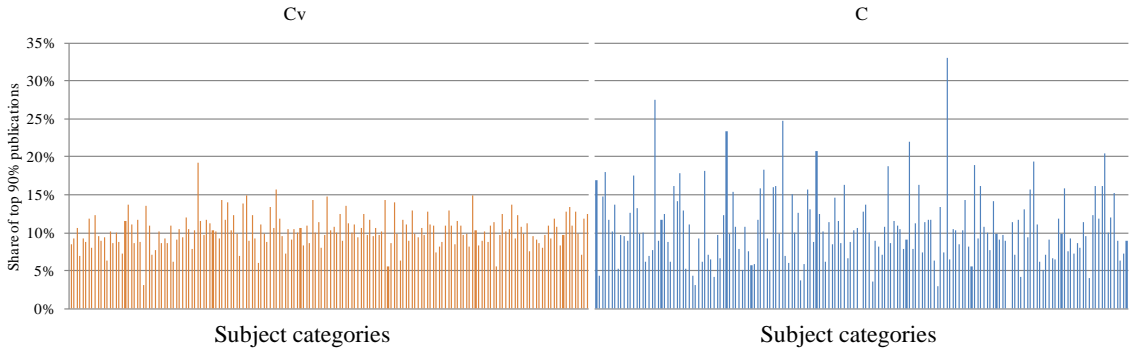

Table 4 provides the descriptive statistics for these analyses extended to all the years: the average shift around the benchmark value of 10% is greater for $C$ than for $C_v$ in almost all years (exception in only 2010). Still, observing the min-max variability, we can clearly see that using $C$ leads to distributions of highly-cited papers that are highly polarized, in the way of SCs without top papers and others where the share of top papers arrives at the extreme of almost half of total (47.9% in 2005). Overall, the variability of the shifts measured by standard deviation is always greater for $C$ than for $C_v$, and almost always double.

*Table 4: Descriptive statistics of share of top 10% cited publications among subject categories*

| | $C$ | | | | $C_v$ | | | |
|---|---|---|---|---|---|---|---|---|
| Year | Average | Min | Max | St. dev. | Average | Min | Max | St. dev. |
| 2004 | 10.7% | 0.8% | 44.4% | 0.053 | 10.2% | 4.1% | 19.0% | 0.019 |
| 2005 | 10.5% | 1.8% | 47.9% | 0.052 | 10.0% | 4.3% | 15.1% | 0.021 |
| 2006 | 10.8% | 0.0% | 33.3% | 0.051 | 10.1% | 4.6% | 15.2% | 0.019 |
| 2007 | 10.3% | 0.6% | 35.4% | 0.044 | 9.9% | 1.9% | 14.7% | 0.019 |
| 2008 | 10.1% | 0.0% | 32.9% | 0.044 | 10.0% | 3.1% | 15.0% | 0.019 |
| 2009 | 10.1% | 0.0% | 26.3% | 0.042 | 10.1% | 4.0% | 16.1% | 0.019 |
| 2010 | 10.1% | 0.0% | 21.6% | 0.042 | 9.7% | 1.6% | 17.9% | 0.021 |
| 2011 | 9.9% | 0.0% | 29.6% | 0.044 | 9.9% | 3.3% | 15.2% | 0.021 |
| 2012 | 9.7% | 1.9% | 20.9% | 0.039 | 10.0% | 4.7% | 14.4% | 0.018 |

## 5. The effects of releasing the constraint that one citation cannot be worth more than two

In this section, we assess the effects of releasing the constraint that no individual citation (even one from a highly cited paper) be allowed to be worth above two, which is the value of two citations from uncited publications.

We do so by introducing a parameter α > 0 in formula 2, as follows:

$$C_v^* = N + \alpha \sum_{i=1}^{N} e^{\beta \cdot f(c_i)}$$

[6]

By varying the cap parameter $\alpha$, one can decide the extent of the impact of the citing publication to be transferred to the cited one.

In the following, we perform a sensitivity analysis based on three different values of $\alpha$, namely 2, 3 and 5.

As an example, we start with the same four 2011 Italian publications in Engineering, mechanical as indicated in Table 2, receiving 21 citations each. Having received the same number of citations, they all show an identical value of $C$ = 2.974. The value of $C_v$ instead varies (Table 5). Variations differ depending on $\alpha$. In particular, for the first two publications $C_v$ decreases as $\alpha$ increases, while the opposite is true for the other two publications. $C_v$ of the last publication is 1.75 times that of the first when $\alpha$=1, and 2.78 times when $\alpha$=5.

*Table 5: The values of $C_v$ of four 2011 Engineering, mechanical publications, receiving 21 citations each, as a function of $\alpha$*

| WoS code | $C_v$ | | | |
|---|---|---|---|---|
| | $\alpha = 1$ | $\alpha = 2$ | $\alpha = 3$ | $\alpha = 5$ |
| 299562000004 | 2.413 | 2.258 | 2.132 | 1.936 |
| 291316000024 | 2.927 | 2.721 | 2.551 | 2.290 |
| 285726600012 | 3.951 | 4.179 | 4.366 | 4.655 |
| 284970700015 | 4.227 | 4.600 | 4.906 | 5.378 |

The increasing variability of $C_v$, as $\alpha$ increases is also shown in Figure 9 which presents the dispersion plots of $C_v$ vs $C$ for the 95 Italian publications in Transportation in year 2012, per $\alpha$ =1 (left panel) and $\alpha$=5 (right panel). The R-squared value for the linear regression of $C$ vs $C_v$ decreases from 0.925 to 0.832.

*Figure 9: Dispersion of C vs $C_v$ for 95 Italian publications in Transportation, per $\alpha$ = 1 (left panel) and $\alpha$ = 5 (right panel)*

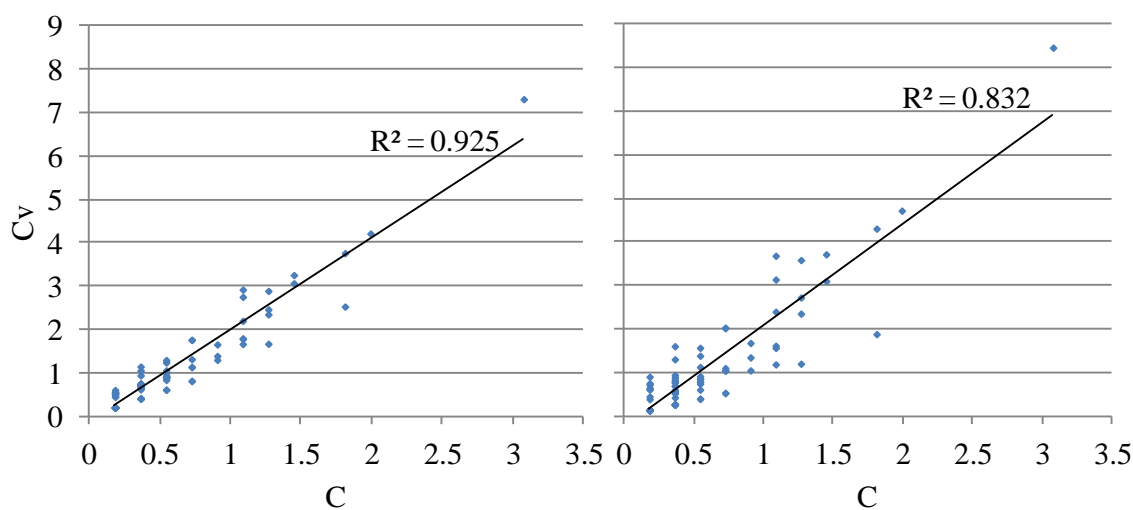

In the case under analysis, a cofounding factor of the lower correlation between $C$ and $C_v$, is the low number of observations (95). When passing to larger-size SCs, the lower correlation as $\alpha$ increases is less noticeable. Table 6 shows the regressions as in Table 3, per each $\alpha$, but for three years only. As $\alpha$ increases the R-squared linear regression of $C$ vs $C_v$ slightly decreases, with values never below 0.95, apart few exceptions, also when α = 5.

*Table 6: R-squared linear regression of $C$ vs $C_v$ for publications in the largest subject categories in each WoS macro-area, for different values of α*

| Year | α | Mathematics | Astronomy & Astrophysics | Chemistry, Multidisciplinary | Environmental Sciences | Biochemistry & Molecular Biology | Oncology | Surgery | Psychology, Experimental | Engineering, Electrical & Electronic | Economics | Political Science |
|---|---|---|---|---|---|---|---|---|---|---|---|---|
| 2004 | 1 | 0.991 | 0.998 | 0.998 | 0.999 | 0.998 | 0.999 | 0.997 | 0.998 | 0.994 | 0.997 | 0.989 |
|  | 2 | 0.985 | 0.996 | 0.996 | 0.997 | 0.997 | 0.999 | 0.995 | 0.995 | 0.990 | 0.995 | 0.982 |
|  | 3 | 0.979 | 0.993 | 0.993 | 0.996 | 0.994 | 0.998 | 0.992 | 0.992 | 0.985 | 0.992 | 0.976 |
|  | 5 | 0.966 | 0.986 | 0.987 | 0.994 | 0.990 | 0.995 | 0.985 | 0.986 | 0.973 | 0.988 | 0.967 |
| 2008 | 1 | 0.987 | 0.997 | 0.998 | 0.995 | 0.998 | 0.999 | 0.995 | 0.996 | 0.990 | 0.993 | 0.977 |
|  | 2 | 0.978 | 0.994 | 0.995 | 0.991 | 0.996 | 0.998 | 0.991 | 0.993 | 0.985 | 0.988 | 0.967 |
|  | 3 | 0.969 | 0.991 | 0.993 | 0.986 | 0.994 | 0.997 | 0.986 | 0.990 | 0.978 | 0.982 | 0.959 |
|  | 5 | 0.952 | 0.983 | 0.987 | 0.977 | 0.989 | 0.995 | 0.977 | 0.983 | 0.964 | 0.973 | 0.945 |
| 2012 | 1 | 0.986 | 1.000 | 0.995 | 0.987 | 0.993 | 0.998 | 0.983 | 0.987 | 0.974 | 0.968 | 0.989 |
|  | 2 | 0.980 | 1.000 | 0.991 | 0.982 | 0.989 | 0.996 | 0.975 | 0.981 | 0.964 | 0.954 | 0.984 |
|  | 3 | 0.974 | 0.999 | 0.985 | 0.977 | 0.984 | 0.995 | 0.966 | 0.976 | 0.952 | 0.941 | 0.980 |
|  | 5 | 0.966 | 0.999 | 0.975 | 0.968 | 0.974 | 0.992 | 0.947 | 0.965 | 0.929 | 0.919 | 0.972 |

Increasing the weight of the citing publications' impact, the variance of $C_v$ increases, and therefore also its ability to reveal significant differences between observations, as compared to $C$. The phenomenon is shown in Figure 10, which reproduces data of Figure 6 for different values of α. In all years but the last, $C_v$ shows a greater capacity in discriminating the differences in impact between publications as α increases.

*Figure 10: Percentage of SCs with variation coefficient for distributions of $C_v$ greater than that for $C$, for different values of α*

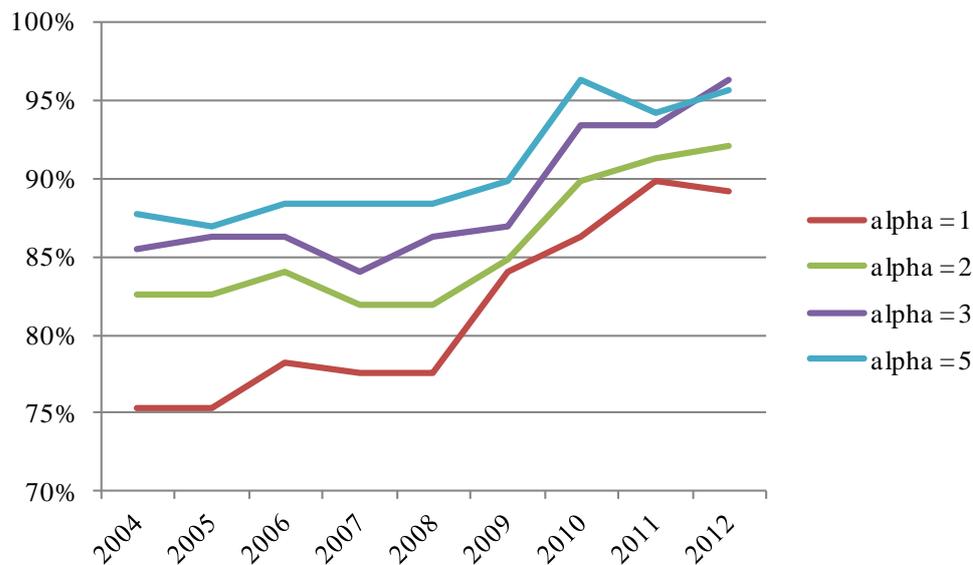

With regard to highly-cited publications, no substantial differences, in terms of polarization in certain SCs, emerge as α increases. This does not mean that α does not affect the status of highly-cited publications, but it does so only slightly. We observe in fact that only 1.8% of highly-cited publications, defined as such for the $C_v$ distribution based on α=1 are not top for α=2, and vice versa. Such percentage raises to 3.4% when contrasting $C_v$ based on α=1 with α=3, and to 5.8% when contrasting $C_v$ for α=1 with α=5.

## 6. Conclusions

The field of bibliometric research has matured remarkably over the course of decades, meaning that scholars now deal with ever more precise questions. However, we remain troubled by a key paradigm, under which *n* citations are always worth more than *n-1*, other conditions (field and citation time window) being equal. The assumption involved is that the citations received by a publication all have the same value, when in reality some of the citing publications would clearly have impacts different than others. The question becomes how such a convention could be set aside, making way for a new paradigm.

The identification of this challenge is not at all new, given that first mention dates back to the 1970s. However the scholars that have taken up the task have offered solutions that are not fully convincing to us, in which the citations received are weighted on the basis of IF, i.e. an indirect measure of their real impact, or on the basis of citations normalized by the length of the relevant reference lists. We therefore propose a new indicator that values citations by the impact of the citing publications, subject to the convention that two citing publications (even if uncited) can never count less than one (even if highly cited).

In the present work we have applied this indicator to the dataset of 2004-2012 Italian WoS publications and carried out comparisons with the traditional indicator based on simple citation counting. The new indicator shows evident advantages, particularly in terms of greater capacity to detect significant differences in impact between publications. Moreover, this greater capacity increases still further with the reduction of the citation time window: a difference that becomes particularly important in applied cases, such as in national research assessment exercises, which for practical reasons generally require a short citation time window. The same advantage would hold true over all citation-based indicators involving a short citation time window, such as those relying on journal impact factor and the like.

The high level of correlation between the distributions may in part be due to the application of the method to the first two citing levels, but also to the weight cap adopted. It attests in every case to convergence between the measures as obtained under the two indicators. This does not imply that the new indicator is superfluous. In absence of an absolute benchmark defining a real measure of impact, where the necessary data are available at a reasonable cost, the more complex "proxy" might be considered. Moreover, the shifts observed between the two measures are frequent and the number of outliers not at all negligible. An observation of particular importance is that the new indicator seems to show greater "sensitivity" when used in identification of the highly-cited papers.

In terms of future developments, different conventions and different contextual conditions can be explored; interesting above all would be recursive procedures capable of accounting for the overall citation network of publications in the reference time window, thus arriving at results equivalent to the "page ranks" of common search engines. A further area of in-depth methodological analysis would concern the document types of the publications: it could be that different document types would require different valuing or scaling. For example, knowing that reviews are on average more cited than research articles, it would be important to empirically measure the advantages to a publication cited by a review, compared to one that is not.

Finally, according to economic theory, the socio-economic returns on research spending depend not only on the degree of diffusion of the resulting knowledge, but also on the rapidity of diffusion. Therefore, all else equal, a citation accrued at time $t$ should be more valuable than a citation at time $t+1$. Given this, we intend to also explore the possibility of valuing citations in a manner that accounts for the time elapsed between the dates of publication of the citing articles and the cited ones.

**References**

Abramo G., (2018). Revisiting the scientometric conceptualization of impact and its measurement. *Journal of Informetrics*, 12(3), 590-597.
Abramo, G., Cicero, T., D'Angelo, C.A. (2012). Revisiting the scaling of citations for research assessment. *Journal of Informetrics*, 6(4), 470-479.
Abramo, G., Cicero, T., D'Angelo, C.A. (2011). Assessing the varying level of impact measurement accuracy as a function of the citation window length. *Journal of Informetrics*, 5(4), 659-667.
Abramo, G., D'Angelo, C.A. (2014). How do you define and measure research productivity? *Scientometrics,* 101(2), 1129-1144.
Abramo, G., D'Angelo, C.A. (2016a). A farewell to the MNCS and like size-independent indicators. *Journal of Informetrics*, 10(2), 646-651.
Abramo, G., D'Angelo, C.A. (2016b). A farewell to the MNCS and like size-independent indicators: Rejoinder. *Journal of Informetrics*, 10(2), 679-683.
Abramo, G., D'Angelo, C.A., (2016c). Refrain from adopting the combination of citation and journal metrics to grade publications, as used in the Italian national research assessment exercise (VQR 2011-2014). *Scientometrics*, 109(3), 2053-2065.
Abramo, G., D'Angelo, C.A., Di Costa, F. (2010). Citations versus journal impact factor as proxy of quality: Could the latter ever be preferable? *Scientometrics*, 84(3), 821-833.
Abramo, G., D'Angelo, C.A., Felici G. (2019). Predicting long-term citations by a combination of early citations and journal impact factor. *Journal of Informetrics,* 13(1), 32-49
Bergstrom, C. T. (2007). Eigenfactor: Measuring the value and prestige of scholarly journals. *College & Research Libraries News*, 68(5), 314-316.
Bollen J., Rodriguez M. A., & van de Sompel, H. (2006). Journal Status. *Scientometrics,* 69(3), 669-687.
Chen, P., Xie, H., Maslov, S., & Redner, S. (2007). Finding scientific gems with Google's PageRank algorithm. *Journal of Informetrics*, 1(1), 8-15.
Cronin, B. (1984). *The citation process: The role and significance of citations in scientific communication*. London: Taylor Graham. ISBN 0-947568-0-1-8

Davis, P.M. (2008). Eigenfactor: Does the principle of repeated improvement result in better estimates than raw citation counts? *Journal of the American Society for Information Science and Technology*, 59(13), 2186–2188.

Ding, Y. (2011). Applying weighted PageRank to author citation networks. *Journal of the American Society for Information Science and Technology*, 62(2), 236-245.

Ding, Y., Yan, E., Frazho, A., & Caverlee, J. (2009). PageRank for ranking authors in co-citation networks. *Journal of the American Society for Information Science and Technology*, 60(11), 2229-2243.

Fiala, D. (2011). Mining citation information from CiteSeer data. *Scientometrics*, 86(3), 553-562.

Fiala, D. (2012a). Bibliometric analysis of CiteSeer data for countries. *Information Processing and Management*, 48(2), 242-253.

Fiala, D. (2012b). Time-aware PageRank for bibliographic networks. *Journal of Informetrics,* 6(3), 370-388.

Fiala, D. (2013a). From CiteSeer to CiteSeerX: Author rankings based on coauthorship networks. *Journal of Theoretical and Applied Information Technology*, 58(1), 191-204.

Fiala, D. (2013b). Suborganizations of institutions in library and information science journals. *Information*, *4*(4), 351-366. 15

Fiala, D. (2014). Sub-organizations of institutions in computer science journals at the turn of the century. *Malaysian Journal of Library and Information Science*, 19(2), 53-68.

Fiala, D., Rousselot, F., & Ježek, K. (2008). PageRank for bibliographic networks. *Scientometrics*, 76(1), 135-158.

González-Pereira, B., Guerrero-Bote, V. P., & Moya-Anegón, F. (2010). A new approach to the metric of journals scientific prestige: The SJR indicator. *Journal of Informetrics*, 4(3), 379–391.

Guerrero-Bote, V. P., & Moya-Anegón, F. (2012). A further step forward in measuring journals' scientific prestige: The SJR2 indicator. *Journal of Informetrics,* 6(4), 674-688.

Kalaitzidakis P., Mamuneas T. P., & Stengos T. (2011). An updated ranking of academic journals in economics. *Canadian Journal of Economics/Revue canadienne d'Economique*, 44(4), 1525-1538.

Kalaitzidakis, P., Stegnos T., and Mamuneas, T.P. (2003). Rankings of academic Journals and institutions in economics, *Journal of the European Economic Association*, 1(6), 1346-1366.

Kochen, M. (1974). *Principles of Information Retrieval*. Wiley, New York. ISBN 9780471496977

Laband, D., and Piette, M. (1994). The relative impact of economic journals. *Journal of Economic Literature*, 32(1), 640-66.

Levitt, J. M., Thelwall, M. (2011). A combined bibliometric indicator to predict article impact. *Information Processing and Management,* 47(2), 300-308.

Li, J., & Willett, P. (2009). ArticleRank: A PageRank-based alternative to numbers of citations for analysing citation networks. *Aslib Proceedings*, *61*(6), 605–618.

Liebowitz, S.J., & Palmer, J.P. (1984). Assessing the relative impacts of economics journals. *Journal of Economic Literature,* 22(1), 77-88.

Ma, N., Guan, J., & Zhao, Y. (2008). Bringing PageRank to the citation analysis. *Information Processing and Management*, 44(2), 800-810.

Nykl, M., Ježek, K., Fiala, D., & Dostal, M. (2014). PageRank variants in the evaluation of citation networks. *Journal of Informetrics*, 8(3), 683-692.

Pinski, G., & Narin, F. (1976). Citation influence for journal aggregates of scientific publications: Theory, with application to the literature of physics. *Information Processing and Management,* 12(5), 297-312.

Radicchi, F., Fortunato, S., Markines, B., & Vespignani, A. (2009). Diffusion of scientific credits and the ranking of scientists. *Physical Review E*, *80*(5), art. no. 056103

Stern D.I., (2014). High-ranked social science journal articles can be identified from early citation information. *PLoS ONE*, 9(11), 1-11.

Su, C., Pan, Y.T., Zhen, Y.N., Ma, Z., Yuan, J.P., Guo, H., …, Wu, Y.S. (2011). PrestigeRank: A new evaluation method for papers and journals. *Journal of Informetrics*, *5*(1), 1–13.

Waltman, L., van Eck, N.J., van Leeuwen, T.N., Visser, M.S., & van Raan, A.F.J. (2011). Towards a new crown indicator: Some theoretical considerations. *Journal of Informetrics*, 5(1), 37-47.

Waltman, L., & Yan, E. (2014). PageRank-Related Methods for Analyzing Citation Networks. In Ding, Y., Rousseau, R. & Wolfram, D. (Eds), *Measuring Scholarly Impact: Methods and Practice* (pp. 83-100). *Springer International Publishing*.

Walker, D., Xie, H., Yan, K.-K., & Maslov, S. (2007). Ranking scientific publications using a model of network traffic. *Journal of Statistical Mechanics: Theory and Experiment*, 6, art. no. P06010.

West J.D., Bergstrom T.C., & Bergstrom C.T. (2010). The Eigenfactor[TM] Metrics: A network approach to assessing scholarly journals. *College of Research Libraries*, 71(3), 236-244.

West, J. D., Jensen, M. C., Dandrea, R. J., Gordon, G. J., & Bergstrom, C. T. (2013). Author-level Eigenfactor metrics: Evaluating the influence of authors, institutions, and countries within the social science research network community. *Journal of the American Society for Information Science and Technology,* 64(4), 787-801.

Yan, E., Ding, Y., & Sugimoto, C. R. (2011). P-rank: An indicator measuring prestige in heterogeneous scholarly networks. *Journal of the American Society for Information Science and Technology*, 62(3), 467-477.

Yan, E. (2014). Topic-based PageRank: Toward a topic-level scientific evaluation. *Scientometrics*, 100(2), 407-437.

Yan, E., & Ding, Y. (2010). Weighted citation: An indicator of an article's prestige. *Journal of the American Society for Information Science and Technology*, 61(8), 1635-1643.

Yan, E., & Ding, Y. (2011). Discovering author impact: A PageRank perspective. *Information Processing and Management*, 47(1), 125-134.